\documentclass[superscriptaddress,showpacs,aps,twocolumn,prb,floatfix]{revtex4-1}
\usepackage{amssymb}
\usepackage{amsmath}
\usepackage[dvips]{graphicx}
\usepackage{bm}
\usepackage{times}

\newcommand{\smeq}{\! = \!}

\newcommand{\smmi}{\! - \!}

\newcommand{\Ef}{E_{\text{F}}}

\newcommand{\bea}{\begin{eqnarray}}
\newcommand{\eea}{\end{eqnarray}}

\newcommand{\ci}{\mathrm{i}}

\newcommand{\ket}[1]{| #1 \rangle}

\begin{document}

\title{Magnetic Structure of Hydrogen Induced Defects on Graphene}
\author{J. O. Sofo}
\affiliation{Physics Department, Pennsylvania State University, University Park, PA
16802,USA}
\author{Gonzalo Usaj}
\affiliation{Centro At{\'{o}}mico Bariloche and Instituto Balseiro, CNEA, 8400 Bariloche,
and CONICET, Argentina}
\author{P. S. Cornaglia}
\affiliation{Centro At{\'{o}}mico Bariloche and Instituto Balseiro, CNEA, 8400 Bariloche,
and CONICET, Argentina}
\author{A. M. Suarez}
\affiliation{Physics Department, Pennsylvania State University, University Park, PA
16802,USA}
\author{A. D. Hern\'andez-Nieves}
\affiliation{Centro At{\'{o}}mico Bariloche and Instituto Balseiro, CNEA, 8400 Bariloche,
and CONICET, Argentina}
\author{C. A. Balseiro}
\affiliation{Centro At{\'{o}}mico Bariloche and Instituto Balseiro, CNEA, 8400 Bariloche,
and CONICET, Argentina}
\date{submitted, December 2011}

\begin{abstract}
Using density functional theory (DFT), Hartree-Fock, exact diagonalization, and numerical renormalization group methods we study the electronic structure of diluted hydrogen atoms chemisorbed on graphene. 
A comparison between DFT and Hartree-Fock calculations allows us to identify the main characteristics of the magnetic structure of the defect. We use this information to formulate an Anderson-Hubbard model that captures the main physical ingredients of the system, while still allowing a rigorous treatment of the electronic correlations.
We find that the large hydrogen-carbon hybridization puts the structure of the defect half-way between the one corresponding to an adatom weakly coupled to pristine graphene and a carbon vacancy. 
The impurity's magnetic moment leaks into the graphene layer where the electronic correlations on the C atoms play an important role in stabilizing the magnetic solution. Finally, we discuss the implications for the Kondo effect.
\end{abstract}

\maketitle

\section{Introduction}
The unusual electronic and mechanical properties of graphene\cite%
{Novoselov2005,Katsnelson2006,Novoselov2007,Geim2007,CastroNeto-review,DasSarma2011}
 triggered an intense research activity after the first isolation of individual graphene sheets.\cite%
{Novoselov2004,Novoselov2005,Zhang2005}  The anomalous charge transport properties of pure
graphene are a consequence of its particular sub-lattice symmetry that leads
to a band structure with a point-like Fermi surface and a linear dispersion
close to the Fermi energy. \cite{Wallace1947}
These properties are responsible
for the peculiar low energy excitations, that correspond to massless-Dirac fermions, and for the long electron mean-free path. 

The possibility of doping graphene
with electrons or holes using external gates, which does not lead to a significant
loss of mobility, allows the design of graphene-based devices.
Another approach, that is also viewed as a promising way to engineer the band structure and  modify its electronic properties, is the controlled adsorption of different types of atoms or molecules on graphene.\cite{Lehtinen2003,Duplock2004,Meyer2008,Chan2008,Uchoa2008,CastroNeto2009,Boukhvalov2008,Boukhvalov2009,CastroNeto2009a,Cornaglia2009,Wehling2009,Wehling2010,Wehling2010a,Ao2010,HernandezNieves2010,Chan2011} 
In particular, there has been an increasing interest to incorporate magnetic effects in
graphene to use graphene-based devices in spintronics.\cite{Tombros2007,Yazyev2008,Wimmer2008,Uchoa2008,Cornaglia2009,Usaj2009,Soriano2010} There are already some advances in this direction, like the injection of spin
polarized currents from ferromagnetic electrodes.\cite{Tombros2007,Jozsa2009,Han2010} An alternative route is the generation of magnetic defects using, for example, adatoms, vacancies or
edges. Hydrogen (H) impurities and carbon (C) vacancies are among the
simplest and most studied point-like magnetic defects.\cite{Yazyev2007,Palacios2008,Yazyev2008b,Balog2009,Yazyev2010,Haase2011} 

The problem of H
atoms chemisorbed on graphene has been studied by several groups.\cite{Duplock2004,Boukhvalov2008,Boukhvalov2009,Casolo2009,Elias2009,Soriano2010,Ao2010} 
It is
known that H impurities are adsorbed on top of a C atom, forming a covalent
bond and locally distorting the honeycomb lattice. There is an
increasing consensus that H adatoms acquire a magnetic moment, therefore
behaving as magnetic impurities. A simple picture used to describe H on
graphene is based on the Anderson model where the localized $1s$ H
orbital is hybridized with the graphene extended states. The intra-atomic
Coulomb repulsion at the $1s$-orbital, together with the graphene pseudo-gap
favors the existence of a local moment at the H impurity.\cite{Uchoa2008,Cornaglia2009} 

The case of C vacancies is quite different. In the ideal case where the lattice remains undistorted after the removal of a C atom,\cite{Yazyev2008} there is a reduction of the coordination number of the three neighboring atoms that generates a
resonant state at the pseudo-gap.\cite{Pereira2006} The existence of this resonance together
with the intra-atomic Coulomb repulsion at the C orbitals naturally leads to
the formation of a magnetic moment. This simple picture is consistent with
the observation of magnetic, Kondo-like correlations in irradiated
samples\cite{Chen2011} and enlightens the role of the on-site Coulomb interaction in
graphene (from hereon denoted as $U_{C}$). The parameter $U_{C}$ has been
estimated to be of the order of a few electronvolts.\footnote{There is, however, some controversy in the literature  regarding wether the  value of $U_C$ is closer to the critical value
that would generate a magnetic instability in pure graphene.\cite{Wehling2011}
Within the context of DFT, we used the \textit{Quantum Expresso} package\cite{Cococcioni2005} to obtain $U_C\sim3.2-3.5$eV.} 

A realistic
description of impurities on graphene should then include the effects of the
electronic correlations in both the impurity and the C orbitals by means of
an Anderson-Hubbard like model. Such a model can extrapolate between two
different limits: (\textit{i}) for a small value of the hybridization between the
impurity and C orbitals, a magnetic moment that is mainly localized at the
impurity orbital can be stable; and (\textit{ii}) for a large hybridization, bonding and
anti-bonding states are shifted away of the low energy region. The effect of
these shifts is to effectively remove the $p_{z}$ orbital of the hybridized
carbon atom from the Fermi energy generating a vacancy-like defect with a
magnetic moment localized at the C atoms. 

As we will discuss below in
detail, we find that the case of a H adatom on graphene is half a way between these
two limiting cases. We show that the H impurity creates a defect with a
complex magnetic structure where a spin $S=\frac{1}{2}$ is localized in a linear
combination of the H and the neighboring C orbitals showing some of the properties of the
vacancy-like defect. 

The paper is organized
as follows: In Section \ref{DFT} we revisit the Density Functional Theory
(DFT) for diluted H on graphene and use the results as a guide to estimate
realistic effective parameters. In Section \ref{HF} we present an
Anderson-Hubbard model and its Hartree-Fock solution is compared with 
DFT results. In Section \ref{NRG} we show that the results can be
interpreted in terms of an effective model of a H-C's cluster embedded in
an effective medium and use the Numerical Renormalization Group (NRG) to
analyze the stability of the magnetic solution.
Finally, a summary and the conclusions are presented in Section \ref{conclu}.

\section{DFT results\label{DFT}}

We consider a graphene layer with a low concentration of H atoms. We
calculate the spin dependent electronic structure of systems with a unit
cell of $N$ C atoms and one H atom where $N$ ranges between $32$
and $72$. Our DFT calculations were done with a plane wave basis as
implemented in the VASP code \cite {Kresse1996a,*Kresse1996} and the cutoff energy
was set to $400$eV. The core electrons were treated with a frozen core
projector augmented wave (PAW) method.\cite{Blochl1994,Kresse1999} The frozen core pseudo-potential for carbon used a core radius for the $s$ partial waves of  $1.20$ a.u. and $1.5$ a.u. for the $p$ channel. All partial waves for H where treated with a cutoff radius of $1.10$ a.u..
We use the PBE generalized gradient approximation to treat the exchange and
correlations.\cite{Perdew1996,Perdew1997} To correct for the dipole moment
generated in the cell and to improve convergence with respect to the
periodic cell size, monopole and dipole corrections were considered.\cite{Neugebauer1992,*Makov1995} 
\begin{figure}[t]
\includegraphics[width=0.45\textwidth]{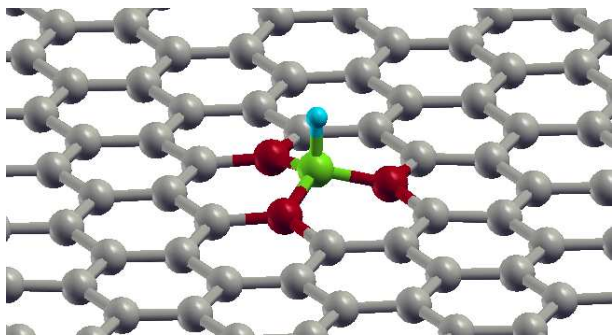}
\caption{(Color online) 3D view of a supercell containing 72C atoms (large spheres) and one H impurity (small sphere) on top of a C atom (C$_{0}$) represented by a green (light gray) sphere. The second nearest-neighbors C atoms (C$_{n}$ with $n=1,2,3$) are represented by red (dark gray) spheres. }
\label{scheme-H}
\end{figure}

Figure~\ref{scheme-H} shows the relaxed structure of a supercell of $72$ C atoms containing one H adatom.  After minimizing the total energy with respect to the coordinates of all atoms in the unit
cell, we find a H induced distortion consisting of a puckering of the
hybridized carbon atom. The distortion is due to the modification of the
electronic structure of the C atom bonded with the H adatom that changes from an $sp^{2}$ configuration  to an $sp^{3}$-like configuration after the hybridization with the
H orbital. In the following, we use the label C$_{0}$ for the C atom directly bonded to the H impurity. C$_{0}$ is represented by a green (light gray) sphere in Fig.~\ref{scheme-H}. The second nearest-neighbors C atoms are represented by red (dark gray) spheres and are labeled as C$_{n}$ with $n=1,2,3$.

\begin{figure}[t]
\includegraphics[width=0.4\textwidth]{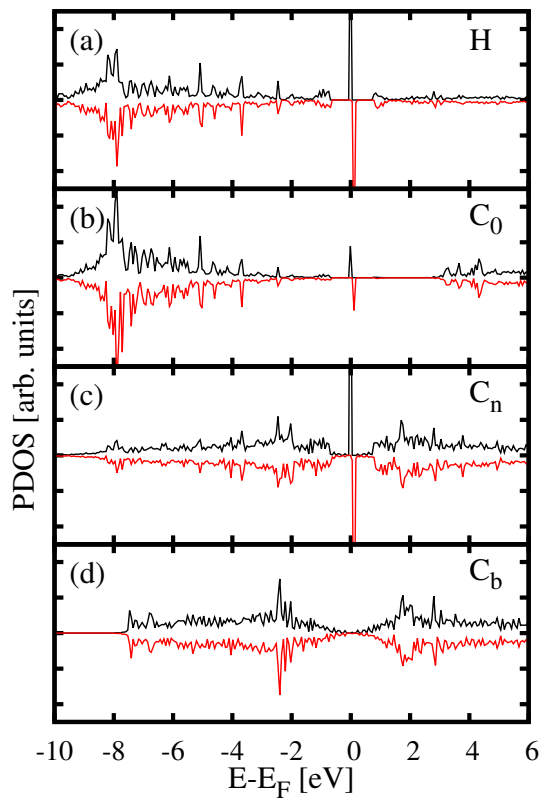}
\caption{(Color online) Partial density of states (PDOS) for the undoped case projected on (a) the H atom, and the $p_z$-orbitals of (b) the nearest-neighbor carbon atom (C$_{0}$), (c) one of the second nearest-neighbors carbon atoms (C$_{n}$), and (d) for a C atom located far from the adatom. For each case case both the majority and minority spin PDOS are plotted.  A clear spin-polarized resonant state near the Fermi level $\Ef$ is observed in Figs.~\ref{DOS-pz}(a) and \ref{DOS-pz}(c).}
\label{DOS-pz}
\end{figure}

Partial densities of states (PDOS) are calculated using a projector scheme provided in the PAW implementation of VASP4.6 when LORBIT=11. The PDOS corresponding to the H atom shows the 
existence of a spin-split resonant state close to the Fermi level, see Fig.~\ref{DOS-pz}(a). This induces a total magnetic moment of $\approx$ 0.49 $\mu_{B}$/cell localized around the impurity. 
The covalent bond with the H induces a spin-polarization in the surrounding C atoms. The spin-polarization is small in the nearest C atom (C$_0$), as shown in Fig.~\ref{DOS-pz}(b) where there is only a small spin-polarized peak close to the Fermi level. The magnetic moment is larger at the  second nearest-neighbor C atoms (C$_{n}$) as can be appreciated in the strong spin-polarization of the $p_z$ orbitals of these atoms, see Fig.~\ref{DOS-pz}(c). The spin-polarization induced by the H adatom on the graphene layer tends to decrease at large distances from the absorption point, and far from the H adatom the PDOS corresponding to the $p_z$ orbitals of the C atoms [Fig.~\ref{DOS-pz}(d)] shows a Dirac point similar to the behavior of the C atoms in pristine graphene.

\begin{figure}[tb]
\includegraphics[width=0.45\textwidth]{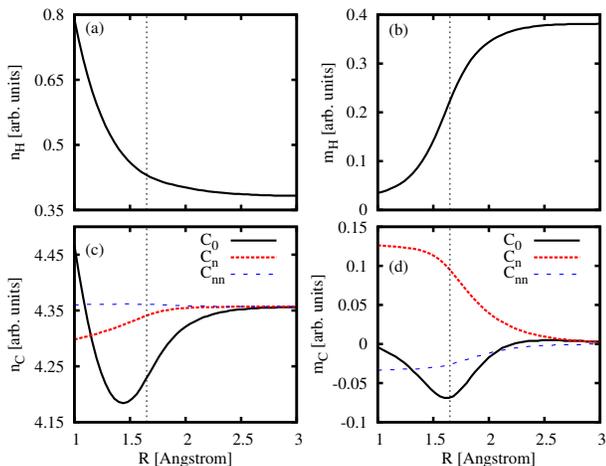}
\caption{(Color online) Charge and magnetization of the most relevant atoms as a function of the distance $R$ between the H atom and the graphene sheet. The equilibrium position of the adatom is indicated by a vertical dashed line.
Notice how the charge (magnetization) of the H atom increases (decreases) as it approaches the graphene sheet. The opposite occurs for the C$_n$ atoms (red short-dashed lines). Both the charge and the magnetization of the C$_0$ atom have a minimum close to the equilibrium position.
}
 \label{Vs-distance}
\end{figure}

The H adatom induces a magnetic moment and charge re-distribution on the surrounding carbon atoms that can be analyzed by monitoring the atomic charge and magnetization as a function of the distance between the H atom and  the graphene plane ($R$). This is shown in Fig.~\ref{Vs-distance}. The vertical dashed line marks the equilibrium position.
The charge of the H atom increases, and its magnetic moment decreases, as it approaches the graphene sheet as shown in Figs.~\ref{Vs-distance}(a) and \ref{Vs-distance}(b). The opposite occurs for the second nearest-neighbors C$_n$ atoms (red short-dashed lines) in Figs. \ref{Vs-distance}(c) and \ref{Vs-distance}(d), i.e., the charge of the C$_n$ atoms decreases while the magnetization increases as the H atom approaches the graphene sheet.  A different behavior can be observed at the absorption site, the C$_0$ atom. In this case, the charge shows a minimum and the absolute value of the magnetization a maximum close to the equilibrium position. 

Far from the absorption site, the perturbation induced by the H atom is smaller and alternates its magnitude between sub-lattices---it is larger on the sub-lattice corresponding to the C$_n$ atoms. For example, the charge of the third nearest-neighbors [C$_{nn}$ atoms in Fig.~\ref{Vs-distance}(c)] is almost independent of the position of the adatom, whereas the magnetization of these atoms only changes slightly compared with the changes in the magnetization of the C$_n$ atoms [compare the red short-dashed and blue dashed lines in Fig.~\ref{Vs-distance}(d)]. This shows that the H atom generates a charge redistribution mostly on the same sub-lattice of the absortion site and Êpredominantly
at the neighboring C atoms, those that were highlighted  with different colors in Fig.~\ref{scheme-H}. As we will see in the next sections, the behavior of the system can be described by a simple model that takes into account the \textit{e-e} correlations only in these atoms and connects them with the rest of the graphene lattice described with a simpler approximation. 

\begin{figure}[t]
\includegraphics[width=0.4\textwidth]{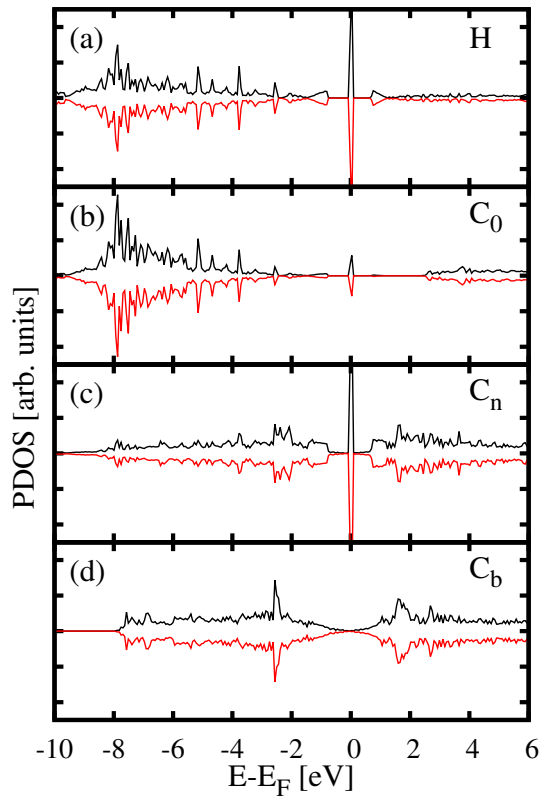}
\caption{(color online) Same as Fig. \ref{DOS-pz} but doped with an extra electron in the unit cell.  Notice the absence of spin-polarization in the system after electron doping.}
\label{DOS+1}
\end{figure}

As was found recently for the case of fluorine adatoms, \cite{Sofo2011} the nature of the chemical bonding of adatoms on graphene can change strongly with electron and hole doping. When graphene is electron-doped, the C$_0$ atom bonded to the F atom retracts back to the graphene plane and for high doping its electronic structure corresponds to nearly a pure $sp^2$ configuration, see Ref.~[\onlinecite{Sofo2011}] for details. The situation is different for the electron doping of graphene with H impurities.
To simulate electron doping we add one electron per unit cell. The extra charge is compensated with a uniform charge background. The results obtained are shown in Fig.~\ref{DOS+1}. The most important difference between Figs.~\ref{DOS-pz} and \ref{DOS+1} is the absence of spin-polarization in the system after electron doping. As we can see, a similar electronic structure with well defined peaks at the Fermi level is observed but neither the H atom, nor the C$_0$ or C$_n$ atoms show spin-polarization---see Figs.~\ref{DOS+1}(a)-\ref{DOS+1}(c). This dependence between gate doping and magnetic moment highlights the delicate interplay between electron correlations and localization in graphene with chemisorbed adatoms. In the following, we use our DFT results as a guide to formulate a theory that goes beyond the mean field level. 

\section{Anderson-Hubbard model\label{HF}}

In order to interpret the DFT results and give a more physical picture of the
magnetic structure of the defect we use a simple Anderson-Hubbard (AH) model given by%
\begin{equation}
\hat{\mathcal{H}}=\hat{\mathcal{H}}_{\mathrm{graph}}+\hat{\mathcal{H}}_{%
\mathrm{imp}}+\hat{\mathcal{H}}_{\mathrm{hyb}}\,,
\end{equation}%
where $\hat{\mathcal{H}}_{\mathrm{graph}}\!$ describes the $\pi $-bands of
the graphene sheet,
\begin{equation}
\hat{\mathcal{H}}_{\mathrm{graph}}\!=\!-\sum_{\langle i,j\rangle \sigma
}t_{ij}\,c_{i\sigma }^{\dag }c_{j\sigma }^{{}}+U_{C}\sum_{i}\,\left( \hat{n}%
_{i\uparrow }\smmi\frac{1}{2}\right) \left( \hat{n}_{i\downarrow }\smmi\frac{1}{2}%
\right) ,
\end{equation}%
here $c_{i\sigma }^{\dag }$ creates an electron with spin $\sigma $ at site 
$i$ of the graphene lattice, the first sum runs over nearest neighbors, 
$U_{C}$ is the Coulomb repulsion in the carbon atoms and $\hat{n}_{i\sigma
}^{{}}=c_{i\sigma }^{\dagger}c_{i\sigma }^{}$ is the number operator of site $i$. The H impurity, which is
bounded to the C$_{0}$ atom located at site $i=0$, is described by 
\begin{equation}
\hat{\mathcal{H}}_{\mathrm{imp}}=\sum_{\sigma }\varepsilon _{H}\,h_{\sigma
}^{\dag }h_{\sigma }^{{}}+U_{H}\,h_{\uparrow }^{\dag }h_{\uparrow
}^{{}}h_{\downarrow }^{\dag }h_{\downarrow }^{{}}
\end{equation}%
where $h_{\sigma }^{\dag }$ creates an electron with spin $\sigma $ at the $%
1s$ orbital of the H impurity with energy $\varepsilon _{H}$ and intra-atomic Coulomb repulsion $U_{H}$. The impurity-graphene interaction includes a
one-body hybridization $V$ and a distortion-induced shift of the C$_{0}$
carbon energy $\varepsilon _{0}$,
\begin{equation}
\hat{\mathcal{H}}_{\mathrm{hyb}}=\varepsilon _{0}\,\hat{n}%
_{0}-V\,\sum_{\sigma }(c_{0\sigma }^{\dag }h_{\sigma }^{{}}\!+\!h_{\sigma
}^{\dag }c_{0\sigma }^{{}})  \notag
\end{equation}%
with $\hat{n}_{0}=c_{0\uparrow }^{\dag }c_{0\uparrow
}^{{}}\!+\!c_{0\downarrow }^{\dag }c_{0\downarrow }^{{}}$. 
In what follows
we assume that the hopping matrix elements $t_{ij}$ with $i,j\neq 0$ are all
equal to $t=2.8$ eV while $t_{0j}=t_{j0}$ is reduced by the distortion.
The energy of the hybridized carbon orbital is given by $\varepsilon
_{0}=A^{2}(\varepsilon _{s}-\varepsilon _{p})$, where $%
\varepsilon _{s}\sim -8$ eV and $\varepsilon _{p}\equiv 0$ are the energies
of the $s$ and $p$ carbon orbitals, respectively, while $A$ is a constant 
 that parametrizes the deformation of the $sp^{2}$
bonding of C$_{0}$.\cite{CastroNeto2009} This expression interpolates between $%
A\!=\!0$ for the $sp^{2}$ and $A\!=\!\frac{1}{2}$  for the $sp^{3}$ configurations. From the DFT
results for the PDOS of a H far from the graphene surface we obtain that $%
\varepsilon _{H}\!\sim \!-1.4t$ and $U_{H}\!\sim \!2.8t$ while we set $%
U_{C}=t$.
Guided by the DFT results we take $t_{0j}=t_{0}=0.6t$, $\varepsilon _{0}=-0.7t$ and $V=2t$. 

\begin{figure}[tb]
\includegraphics[width=0.45\textwidth]{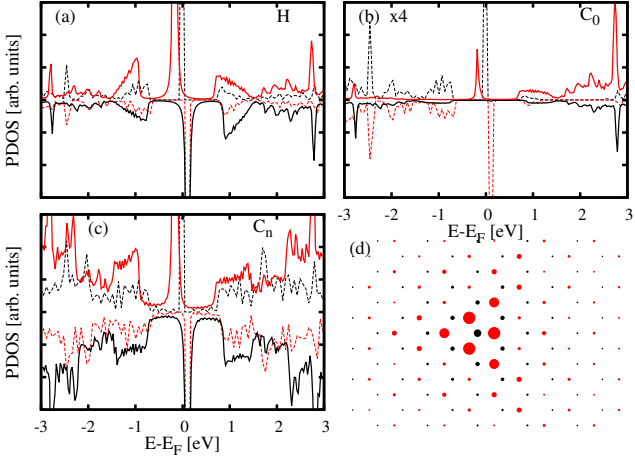}
\caption{(color online) Comparison between the DFT PDOS  (dashed line) projected onto the $p_z$ orbitals and the Hartree-Fock (solid line) results obtained with the full AH model. In both cases we use a $72$C  atoms unit cell. The right bottom panel shows the spatial distribution of the magnetization on the C atoms obtained with the AH model for a unit cell of 288 C atoms where finite size effects are negligible.}
\label{DOS-pz-comparison}
\end{figure}

Figure \ref{DOS-pz-comparison} shows a comparison of the DFT results for the
PDOS projected onto the $p_{z}$ orbitals with those obtained with the AH model within the Hartree-Fock (HF) approximation. In both cases we used a cell of $72 $ C atoms to facilitate the comparison. There is a good  qualitative agreement
between both methods, which indicates that the chosen effective AH parameters are adequate for capturing the relevant aspects of the problem---it should be emphasized that we do not intent to fit the parameters but find a reliable range of values for them. 
The overall qualitative agreement include some features related to the finite size of the cell as, for example, the appearance of gap-like and sharp peak structures. This is a consequence of the interference effects introduced by the periodic array of impurities (all in the same sub-lattice). Within our TB model, such an effect can be eliminated, without much numerical effort, by increasing the size of the unit cell. Fig.  \ref{DOS-pz-comparison}(d) shows the spatial profile of the magnetization on a cell containing $288$ C atoms, which shows a triangular symmetry characteristic of an isolated impurity.

\subsection{Minimal Anderson-Hubbard model}

\begin{figure}[tb]
\includegraphics[width=0.45\textwidth]{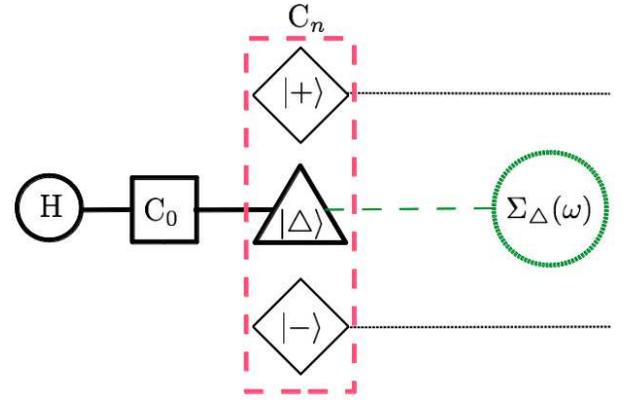}
\caption{Schematic representation of the MAHM. The H impurity, the C$_{0}$
atom and the state $|\triangle \rangle $ are treated exactly while the rest
of the C atoms are treated as a non-interacting reservoir represented by the
self-energy $\Sigma _{\triangle }(\protect\omega )$. The capacitive coupling to the $|\pm\rangle$ states is taken into account only through a mean field approximation (see text).}
\label{scheme}
\end{figure}

The AH model presented above can be further reduced to a much
simpler model that still captures the relevant aspects of the problem, allows to consider a single isolated impurity, greatly reduces the numerical work, and reproduces to an excellent accuracy the results of the full HF approach obtained with large unit cells. We
first note that within the HF approximation, the presence of the H impurity
generates a small charge redistribution mainly on its  neighboring C atoms. Namely, in the C$_{0}$ atom and the three next nearest-neighbors, C$%
_{n}$ with $n=1,2,3$ (see Fig. \ref{scheme-H}). Therefore, it is sufficient to limit ourselves to consider  a small cluster
embedded in an effective medium where the energies of the $p_{z}$ orbitals
are fixed and  the occupation numbers are solved self-consistently in the
cluster. This approximation is numerically simpler as the
self-consistent equations can be expressed in terms of integrals of
analytical functions. Moreover, based on this approximation, the reduced
Hamiltonian can be treated using more powerful numerical tools
like exact diagonalization or the Numerical Renormalization Group (NRG) presented in the next section.

To illustrate the procedure, let us first consider the one-body part of $%
\hat{\mathcal{H}}$. Due to the hexagonal structure of the lattice, the C$%
_{0} $ atom only couples to the symmetric combination of the $p_{z}$
orbitals of its nearest neighbor carbon atoms (C$_{n}$). We denote that state
as $|\triangle \rangle $ and the corresponding fermionic operator as $%
c_{\triangle \sigma }^{\dagger }=\sum_{n}c_{n\sigma }^{\dagger }/\sqrt{3}$
where $n=1,2,3$ labels the C$_{n}$ atoms (we will also refer to this state as C$_\triangle$). 
The other two orthogonal linear
combinations of the C$_{n}$ $p_{z}$-orbitals, denoted by $|\pm \rangle $,
are not directly coupled to C$_{0}$. Furthermore, because of the symmetry of
the hexagonal lattice, the states $|\triangle \rangle $ and $|\pm \rangle $
are not coupled by the rest of the lattice either. As a result, the one-body
terms of $\hat{\mathcal{H}}$ can be separated into three decoupled
parts---this is schematically shown in Fig. \ref{scheme}.

Hence, for calculating
the properties of the H impurity, the C$_{0}$ atom and C$_\triangle$,
it is sufficient to consider a reduced Hamiltonian for the reduced system (see Fig. \ref{scheme})
and include the rest of the lattice as an effective self-energy 
$\Sigma _{\triangle }(\omega )$---the calculation is presented in the
appendix. More explicitly, the one-body Green function of the reduced system
can be written as $\bm{G}(\omega )=[\omega \bm{I}-\bm{\mathcal{H}}%
_{r}(\omega )]^{-1}$ with

\begin{widetext}
\begin{equation}
\bm{\mathcal{H}}_{r}(\omega )=\left( 
\begin{matrix}
\varepsilon _{H} & V & 0 & 0 & 0 \\ 
V & \varepsilon _{0} & \sqrt{3}t_{0} & 0 & 0 \\ 
0 & \sqrt{3}t_{0} & \varepsilon _{\triangle }\!+\!\Sigma _{\triangle
}(\omega ) & 0 & 0 \\ 
0 & 0 & 0 & \varepsilon _{+}\!+\!\Sigma _{+}(\omega ) & 0 \\ 
0 & 0 & 0 & 0 & \varepsilon _{-}\!+\!\Sigma _{-}(\omega )%
\end{matrix}%
\right) \,,  \label{Hr}
\end{equation}%
\end{widetext}
where we have included the possibility that the C$_{n}$ atoms have a
different energy than the rest of the C atoms of the graphene lattice (and included the $|\pm \rangle $ states for completeness). So
far this is an exact procedure. The addition of the Coulomb interactions $%
U_{H}$ and $U_{C}$ \textit{only} in C$_{0}$ can still be treated in a
similar way, provided we use the appropriate method, as the rest of
the system remains a non-interacting fermionic bath and it can still be
represented by $\Sigma _{\triangle }(\omega )$. This is no longer true when
interactions are included in the rest of the C atoms. We will argue,
however, that for a qualitative understanding it is sufficient, and important, to take into account the 
$U_{C}$ interaction \textit{only on} C$_{0}$ and the C$_{n}$ ($n=1,2,3$)
atoms. We will refer to this model as the Minimal Anderson-Hubbard Model
(MAHM).

At the level of the HF approximation, the Coulomb repulsion $U_{C}\sum_{i=1,2,3}\,\left( 
\hat{n}_{i\uparrow }-\frac{1}{2}\right) \left( \hat{n}_{i\downarrow }-\frac{1%
}{2}\right) $ shifts the energy of the states $|\triangle \rangle $ and $%
|\pm \rangle $ \ preserving the form of the effective Hamiltonian $\bm{\mathcal{H}}_{r}(\omega )$
for each spin projection and then the propagators. The spin dependent self-consistent Green function $%
\widetilde{\bm{G}}_{\sigma }$ of the system is obtained from Eq. (\ref{Hr}) with the
self-consistent energies $\tilde{\varepsilon}_{H}^{\sigma }=\varepsilon
_{H}+U_{H}(\langle \hat{n}_{H\bar{\sigma} }\rangle -\frac{1}{2})$, $\tilde{\varepsilon}%
_{0}^{\sigma }=\varepsilon _{0}+U_{C}(\langle \hat{n}_{0\bar{\sigma }}\rangle
-\frac{1}{2})$, and 
\begin{equation}
\tilde{\varepsilon}_{\triangle }^{\sigma }=\varepsilon _{\triangle
}+U_{C}\left( \frac{\langle \hat{n}_{\triangle }^{\bar{\sigma} }\rangle +\langle 
\hat{n}_{+}^{\bar{\sigma }}\rangle +\langle \hat{n}_{-}^{\bar{\sigma }}\rangle }{3}-%
\frac{1}{2}\right) \,
\end{equation}%
with $\bar{\sigma}=-\sigma$.
We have  tested the validity of the MAHM by comparing the PDOS of
the H, the C$_{0}$, and the C$_{n}$ atoms with those obtained with 
the full model which includes the $U_{C}$ interactions everywhere. The agreement between both approaches is excellent, provided the unit cell used in the later case is large enough for the finite cell size effects to be negligible. It is also worthy to emphasize that the MAHM leads to a magnetic structure similar to the one shown in Fig. \ref{DOS-pz-comparison}(d), which indicates that much of the  observed antiferromagnetic structure is related to Friedel oscillations.

\begin{figure}[b]
\includegraphics[width=0.4\textwidth]{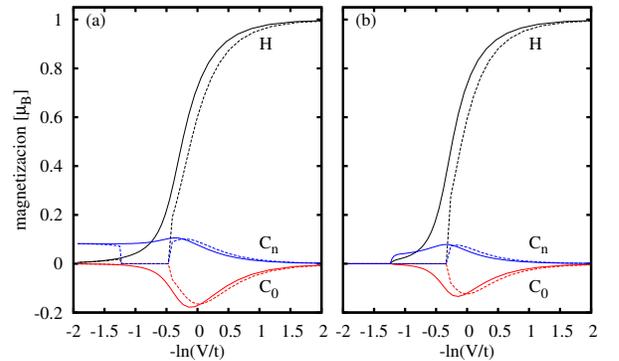}
\caption{(color online) Site magnetization  as a function of the hybridization $V$ between the H and C$_0$ atoms obtained within the MAHM for two different values of $\varepsilon_H$: $\varepsilon_H=1.29t$ (solid lines) and $\varepsilon_H=0.7t$ (dashed lines) and for $U_C=t$ (a) and $U_C=0$ (b).  Notice that the simple model is able to capture the general trend of the spatial distribution of the magnetic structure (compare with Fig. \ref{Vs-distance}). The presence of $U_C$ not only induces some magnetization of the C$_n$ atoms for large $V$ but can also lead to a re-entry behavior of the magnetization.}
 \label{mag-MAHM}
\end{figure}

It is interesting to compare results of the MAHM  with those of DFT. 
For that purpose, we plot in Fig. \ref{mag-MAHM}(a) the magnetization of the H, C$_0$  and C$_n$ atoms  as a function of the  hybridization $V$ for  two different values of $\varepsilon_H$. These results should be compared with those in Fig. \ref{Vs-distance}. We clearly see that the MAHM is able to capture the most relevant features of the DFT results. Namely, that the magnetic state of the impurity is somewhere in between a pure adatom state ($V\ll t$), where the magnetization is mainly localized at the H atom,  and a vacancy state ($V\rightarrow \infty$), where a substantial amount of magnetization has been transferred to the C$_n$ atoms (mainly  dominated by the  $|\triangle \rangle $ state). 
For a quantitative comparison of the results we have to consider
the fact that in the DFT approach the lattice is relaxed for each H-graphene
distance and consequently other parameters, like $\varepsilon _{0}$ and $%
t_{0}$, also depend on the distance.
Figure \ref{mag-MAHM}(b) shows the same parameters but in the absence of the \textit{e-e} interaction on the C atoms. Clearly,  the $U_C$ interaction plays an important role in the case of large $V$ (vacancy-like state), being responsible of the re-entry behavior observed in Fig. \ref{mag-MAHM}(a).

Before discussing the effect of the interactions beyond the HF
approximation, we note that for $\omega \rightarrow 0$ the self-energy $\Sigma _{\triangle }(\omega )\sim
\alpha \omega \ln \left\vert \omega \right\vert -\ci \sqrt{3} \left\vert \omega
\right\vert $, and the spectral density of the $|\triangle \rangle $ state
presents a divergence at the Dirac point, $-\mathrm{Im}(g_\triangle)\smeq =-\mathrm{Im}(1/[\omega-\Sigma_\triangle(\omega)])$. This is precisely the vacancy
state (projected onto the $|\triangle \rangle $ state) that has been
extensively discussed in the literature.\cite{Pereira2006,Peres2006} This singular DOS is what makes the 
$|\triangle \rangle $ state unstable against the formation of a localized
magnetic moment when the interactions are included. 
 Conversely, the $\!\Sigma _{\pm }(\omega )$ self-energies
diverge at the Dirac point and the spectral densities of the $|\pm \rangle $
states show a pseudo-gap  at low energies (see the appendix).
In view of this, one can expect that the main role of the $U_C$ interaction at low energies will manifest  through the weakest coupled state $\ket{\triangle}$.
Therefore, we only keep the  
$U_{C}/3 \left(\hat{n}_{\triangle\uparrow }-\frac{1}{2}\right) \left( 
\hat{n}_{\triangle\downarrow }-\frac{1}{2}\right) $
part of the interaction between the C$_{n}$ ($n=1,2,3$) atoms in what follows and neglect the rest. 

Under these assumptions the reduced Hamiltonian (MAHM) describes a correlated three-site
cluster, given by the H, the C$_{0}$ and the $|\triangle \rangle $ orbitals,
embedded in an effective medium with a pseudo-gap described by
 $\!\Sigma _{\triangle }(\omega )$ --- the mean field charge interaction
between the $|\triangle \rangle $ and the $|\pm \rangle $ orbitals 
that tends to keep the charge neutrality of the $|\triangle \rangle $ state can be included through an effective site energy $\overline{\varepsilon}_\triangle$. In the
next section we analyze the properties of this model.

\section{Exact diagonalization and NRG results\label{NRG}}

The reduced Hamiltonian has the form 

\begin{equation}
\hat{\mathcal{H}}_{R}=\hat{\mathcal{H}}_{\mathrm{cluster}}+\hat{\mathcal{H}}%
_{\mathrm{band}}+\hat{\mathcal{H}}_{\mathrm{hyb}}\,,
\end{equation}%
where

\begin{widetext}
\begin{eqnarray}
\nonumber
\hat{\mathcal{H}}_{\mathrm{cluster}} &=&\varepsilon _{H}\,\hat{n}%
_{H}+U_{H}\,h_{\uparrow }^{\dag }h_{\uparrow }^{{}}h_{\downarrow }^{\dag
}h_{\downarrow }^{{}} +\varepsilon _{0}\,\hat{n}_{0}+U_{C}\,\left( \hat{n}_{0\uparrow }-\frac{1}{%
2}\right) \left( \hat{n}_{0\downarrow }-\frac{1}{2}\right) +\overline{\varepsilon }_{\triangle }\,\hat{n}_{\Delta }+\frac{U_{C}}{3}%
\,\left( \hat{n}_{\Delta \uparrow }-\frac{1}{2}\right) \left( \hat{n}%
_{\Delta \downarrow }-\frac{1}{2}\right)\\
&&-\,V\sum_{\sigma }(c_{0\sigma }^{\dag }h_{\sigma }^{{}}\!+\!h_{\sigma
}^{\dag }c_{0\sigma }^{{}})-\sqrt{3}t_{0}\sum_{\sigma }(c_{\Delta \sigma }^{\dag }\!c_{0\sigma
}^{{}}+\!c_{0\sigma }^{\dag }\!c_{\Delta \sigma }^{{}})
\end{eqnarray}%
\end{widetext}
and $\hat{\mathcal{H}}_{\mathrm{band}}$ and $\hat{\mathcal{H}}_{\mathrm{hyb}%
}\,$describe a band with a pseudo-gap at the Dirac point and the coupling to
the $|\triangle \rangle $ state 
\begin{equation}
\hat{\mathcal{H}}_{\mathrm{band}}=\sum_{\nu \sigma }\varepsilon _{\nu }f_{\nu
\sigma }^{\dag }f_{\nu \sigma }^{{}}
%\hat{\mathcal{H}}_{\mathrm{band}}=\sum_{\nu \sigma }\varepsilon _{\nu }c_{\nu
%\sigma }^{\dag }c_{\nu \sigma }^{{}}
\end{equation}%and 
\begin{equation}
\hat{\mathcal{H}}_{\mathrm{hyb}}=\sum_{\nu \sigma }t_{\nu }(f_{\nu \sigma
}^{\dag }c_{\Delta \sigma }^{{}}+c_{\Delta \sigma }^{\dag }f_{\nu \sigma
}^{{}})\,,
\end{equation}%
that can be rewritten as
\begin{equation}
\hat{\mathcal{H}}_{\mathrm{hyb}}=\sqrt{2} t\sum_{\sigma }(f_{0 \sigma}^{\dag }c_{\Delta \sigma }^{{}}+c_{\Delta \sigma }^{\dag }f_{0 \sigma
}^{{}}).
\end{equation}%
where $f^\dagger_{0\sigma}=\frac{1}{\sqrt{2}t}\sum_\nu t_\nu f_{\nu\sigma}$, creates an electron on a symmetric combination of H's third-nearest-neighbor C atoms.

\begin{figure}[bt]
\includegraphics[width=0.45\textwidth]{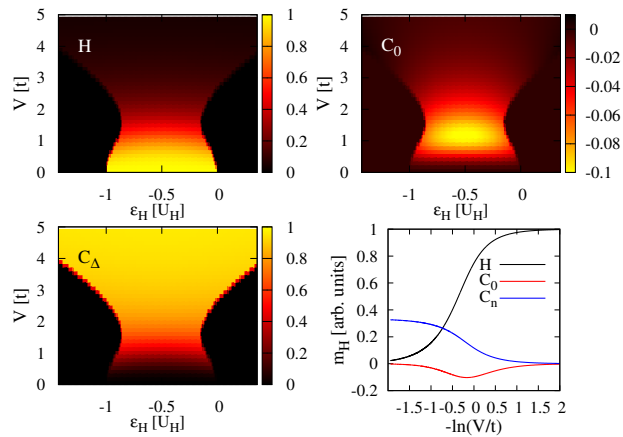}
\caption{(Color online) Color maps showing the magnetization of the H, C $_0$ and C$_\triangle$ atoms as a function of the orbital energy of the H atom, $\varepsilon_H$ and the adatom hybridization $V$ as obtained from the exact diagonalization of $\hat{\mathcal{H}}_{\mathrm{cluster}} $. The right lower panel shows the magnetization  for $\varepsilon_H=-U_H/2$ (note that here the magnetization of C$_n$ is plotted). }
\label{mapa}
\end{figure}

It is instructive to start the analysis by looking at the many-body states
of the isolated cluster. 
For undopped graphene ($\mu =0$), 
we take $\overline{\varepsilon }_{\triangle }=0$, $U_{H}\!\sim \!3t$ 
and $U_{C}=t$ as in the previous sections but take $\varepsilon _{0}=0$ in what follows to simplify the analysis (taking $\varepsilon _{0}\neq0$ introduces a electron-hole asymmetry which is not relevant at this point). 

We diagonalize $\hat{\mathcal{H}}_{\mathrm{cluster}}$ in the different charge
sectors for different values of $\varepsilon _{H}$ and $V$. Figure \ref{mapa} shows
the regions of stability of the different charge states in the $\left[
\varepsilon _{H},V\right] $ plane. For $\varepsilon _{H}$ $\sim -U_{H}\,/2$
the ground state is always a 3-particle state with spin $S=1/2$. The
region of stability of the magnetic states as a function of $\varepsilon _{H}
$ shows a narrowing for $V\sim \!2t$. For $V<\!2t$,  the magnetic moment is
localized mainly at the H orbital while for $V>\!2t$ it is transferred
to the $|\triangle \rangle $ state. Note that, as the hybridization
increases, the spin is transferred directly from the H atom to the $|\triangle
\rangle $ orbitals, in agreement with the DFT and HF results.
This is more clearly seen in Fig. \ref{mapa}(d) where the magnetization of the H, C$_0$ and C$_n$ atoms is plotted as a function of $\ln(V/t)$ for $\varepsilon_H=-U_H/2$ (this is equivalent to plot it as a function of the H\---C$_0$ distance if an exponential dependence of V is assumed).
In agreement with the previous results, the magnetic moment is transferred to the C orbitals as the hybridization increases. 

From these exact results we can now calculate the Kondo coupling constant $J$. This is done in the standard way, eliminating high energy states of the cluster through a Schrieffer--Wolff transformation \cite{Schrieffer1966} to get
\begin{equation} \label{eq:KondoJ}
J=2t^2\sum_{\nu\sigma}\frac{\langle \Uparrow|c^\dagger_{\triangle\sigma}|\nu\rangle\langle\nu|c_{\triangle\sigma}|\Downarrow\rangle+\langle \Uparrow|c_{\triangle\sigma}|\nu\rangle\langle\nu|c^\dagger_{\triangle\sigma}|\Downarrow\rangle}{E_\nu-E_\Uparrow}
\end{equation}
where $|\sigma\rangle$ is the degenerate ground state of the cluster with energy $E_\Uparrow=E_\Downarrow$, and the $|\nu\rangle$ are excited states. 
Note that in the absence of electron-hole symmetry in the cluster there will also be a local potential scattering for the conduction electrons due to the cluster.

\begin{figure}[tb]
\includegraphics[width=0.45\textwidth]{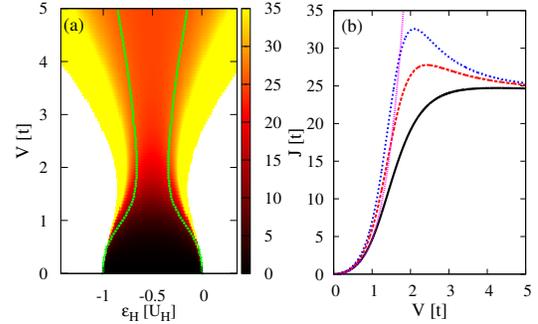}
\caption{(Color online) Kondo coupling constant $J$ estimated from the exact diagonalization of the MAHM cluster. (a) as a function of $\varepsilon_H$ and $V$ and for the same parameters as in Fig. \ref{mapa}. (b) $J$ as a function of V for three different values of $\varepsilon_H$: $-U_H/2$ (solid line), $-0.8t$ (dashed line) and $-0.7t$ (short-dashed line). The case $U_C=0$ (dotted line) is included for comparison.}
\label{J}
\end{figure}

Figure \ref{J}(a) shows the $J$-coupling in the $[\varepsilon_H,V]$ plane while Fig. \ref{J}(b) shows its dependence on $V$ for different values of $\varepsilon_H$. For small $V\ll t$, the magnetic moment, which is mainly localized on the H atom, is weakly coupled to the graphene sheet and the Kondo coupling is small. For a fixed $V$, the Kondo coupling increases as $\varepsilon_H$ approaches the charge degeneracy lines where perturbation theory fails and so $J$ diverges due to vanishing denominators in Eq. (\ref{eq:KondoJ}).
For large $V>2t$ the bond between the H atom and the C$_0$ atom  is strong enough to effectively decouple both atoms from the rest of the system. The magnetic moment is transferred to the $|\triangle\rangle$ state and the Kondo coupling becomes $V$ independent. In the large $V$ limit it is simply given by $J(V\to \infty)= 8t^2/U_\triangle=24 t$.

We complete the analysis by coupling the cluster to the rest of the graphene layer which provides an hybridization $\Gamma_\triangle\sim \sqrt{3}|\omega|$ at low energies. In the parameter region where the cluster is magnetic, we end-up with a pseudo-gap Kondo problem.\cite{Withoff1990,Fritz2004,Cornaglia2009,Jacob2010,Ingersent1996,Vojta2010,Bulla1999,Chen1999,Fritz2006,Vojta2004,Uchoa2011}
In this model, for an electron-hole symmetric impurity, the magnetic
moment of the impurity (the cluster Êin the present case) remains
unscreened even at zero temperature. For the electron-hole asymmetric
situation the magnetic moment can be screened only if the Kondo
coupling is larger than a critical coupling $J_c$ of the order of the
bandwidth.\cite{Withoff1990,Fritz2004} 
We calculated the stability of the magnetic moment in the cluster when it is coupled to the rest of the system. We used the NRG\cite{Wilson1975,Bulla2008} with a pseudo-gapped density of states and a Fermi velocity chosen to reproduce the effect of $\Sigma_\triangle(\omega)$.

The solid line in Fig. \ref{J}(a) shows the NRG results for the stability region of the total magnetic moment on the cluster at zero temperature. We observe that the coupling to the rest of the system reduces the parameter area where the magnetic moment is stable at zero temperature (a similar effect is observed in the Hartree-Fock solution). For $V\lesssim2t$ the magnetic moment is mainly localized on the H atom and it is screened when the Kondo coupling reaches a critical value of $\sim\!10t$ (this unconventional screening is associated to a change of the cluster's charge\cite{Vojta2004}). For larger values of $V$ the magnetic moment is localized on the $|\triangle\rangle$ state.  At finite but large ($>2t$) values of $V$ the critical coupling increases with increasing $V$.
This is due to the fact that, in the present model, if $V\to \infty$ the magnetic impurity becomes electron-hole symmetric for any value of $\varepsilon_H$ and the magnetic moment remains unscreened for any value of $J$. 

It is worth mentioning that in the doped case the magnetization maps of Fig. \ref{mapa} are modified.\cite{Uchoa2008} 
For $V\gtrsim2 t$, the magnetic moment stability region narrows and shifts, being centered around a line given by  $V^2\sim a\, \varepsilon_H+b$. 
This shows that the impurity states is  much more sensitive to doping in  the vacancy-like regime than in the low V regime where the magnetic moment is localized at the H atom. Therefore, a realistic H induced defect on graphene would be more sensitive to doping than what one would expect from the usual Anderson-like model for an impurity on pristine graphene. A detailed  study of this effect will be presented elsewhere.

\section{Summary and Conclusions\label{conclu}}

We have combined a DFT description of diluted H impurities on graphene with
tight-binding and effective models to describe the magnetic structure of a
H induced defect. The DFT approach provides a realistic picture of the
structural distortions around the adsorbed H atom and predicts a magnetic moment
localized in the neighborhood of the adatom. Additionally, it allows us
to estimate parameters that are used to build an effective Anderson-Hubbard
type model Hamiltonian. The model is solved at the mean-field level and the
results are compared with the full DFT band structure to test the quality of
the mapping. The model Hamiltonian is then used to study a \textit{single} H impurity
adsorbed on graphene, a situation that can not be tackled with the current DFT 
based methods and
that allows us to identify in detail the structure of the induced defect as
well as its magnetic properties without the complications generated by the
interactions between impurities.

Within the single impurity Anderson-Hubbard Hamiltonian, the mean field
approximation gives a magnetic solution for undoped graphene and a
strong dependence of the impurity magnetic moment with doping---an
effect with interesting implications for spintronics
and for applications in magneto-transport devices. 

In order to treat the delicate balance between kinetic energy and correlations at the defect 
including quantum
fluctuations, we devised a minimal Anderson-Hubbard model that takes into
account explicitly the electronic correlations at the impurity orbital as
well as on the surrounding carbon atoms and replace the rest of the system
by an effective medium.

An analysis of the isolated cluster 
illustrates the structure of the magnetic moment. For $-U_H<\varepsilon _{H}<0$
and small hybridization $V$ the spin is localized at the H orbital while 
for large $V$ the spin is transferred to the carbon atoms forming a vacancy-like state. For intermediate values of
the hybridization, that correspond to a realistic description of H,
the stability region shows a neck and the magnetic moment is in a linear combination of
the impurity and C orbitals. The effect of the rest of the host
graphene is treated as an effective medium with a pseudo-gap using the NRG.
We consider the case of undoped graphene where the Fermi energy lies at the
Dirac point.  As shown in Fig. \ref{J} the region of stability of the magnetic
moment is narrowed as the cluster is coupled to the rest of the system. 
This behavior can be understood in terms of the known results
for the Anderson impurity model in a system with a graphene-like pseudo-gap:
 in the case of electron-hole symmetry the spin is never
screened,\cite{Fritz2004} while away from the electron-hole symmetry the spin can be screened at
low temperatures if the Kondo coupling is larger than a critical value of
the order of the bandwidth. 
Interestingly, within our model,  in the large $V$ regime (vacancy-like state), the electron-hole symmetry is recovered and the magnetic moment remains unscreened.

While breaking the electron-hole symmetry will modify some of these results (a detailed study will be presented elsewhere),  the main results of the present work are
robust against it: (i)  the spin is transferred to the carbon atoms as the hybridization
increases, (ii) the Kondo coupling can reach quite large values, and (iii) for realistic values of the parameters (obtained from our DFT calculations) the H induced defect is half-way between the one corresponding to an adatom weakly coupled to pristine graphene and a carbon vacancy.

\begin{acknowledgements}
We thank M. Vojta and T. Wehling for useful conversations. JOS and AS acknowledge support from the Donors of the American Chemical Society Petroleum Research Fund and use of facilities at the Penn State Materials Simulation Center. GU, PSC, ADH, and CAB acknowledge financial support from PICTs 06-483 and 2008-2236 from ANPCyT and PIP 11220080101821 from CONICET, Argentina. 
\end{acknowledgements}
\appendix

\section{Non-interacting self-energy}

In Section \ref{HF} we solved the adatom problem using a simplified model where
Coulomb interaction was assumed to be relevant only on the $C_{0}$ and $%
C_{n} $ carbon atoms. In that case, and if we are only interested on the
properties of the reduced system formed by the adatom and the above
mentioned $C$ atoms, the presence of the rest of the graphene sheet can be
taken into account through a self-energy contribution. In the following, we calculate this self-energy using the
Dyson equation and taking advantage of the following: (i) the structure of
the honeycomb lattice around a given atom has the same structure of a Bethe
lattice (up to the second nearest-neighbors) which allows a simple disentanglement of
the lattice Green function; (ii) the exact lattice Green function of a given
site, $\mathcal{G}_{i}$ of the honeycomb lattice has an analytic closed form.\cite%
{Horiguchi1972}

\begin{figure}[tb] %  figure placement: here, top, bottom, or page
   \centering
   \includegraphics[width=.48\textwidth,bb= 18 32 403 301]{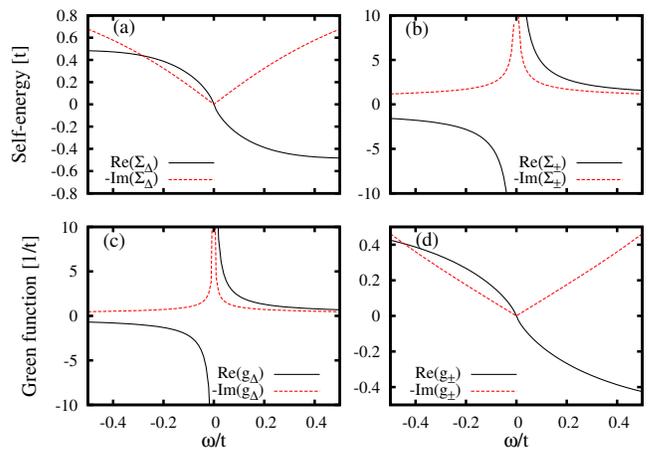} 
   \caption{Self-energy and unperturbed green functions of the C$_n$ atoms. (a) and (c) correspond to the state $|\triangle\rangle$ while (b) and (d) correspond to the $|\pm\rangle$ states.
 Note that $\Sigma_\triangle(\omega)$ presents a pseudo-gap close to the Dirac Point, and then $g_\triangle(\omega)$ shows a resonant-like behavior (vacancy state). The opposite behavior is observed for $\Sigma_\pm(\omega)$ and $g_\pm(\omega)$. }
   \label{sigmas}
\end{figure}

Let us denote by $c_{\triangle }^{\dagger }=\sum_{i}c_{i}^{\dagger
}/\sqrt{3}$ the fermionic operator that creates an electron in an state that
is a symmetric linear combination of the $p_{z}$ orbitals of the three
nearest-neighbors atoms of $C_{0}$. The quantity of interest is the
non-interacting self-energy of that state $\Sigma _{\triangle }(\omega )$
due to the rest of the graphene (without $C_{0}$). That is, if $g_{\triangle
}(\omega )=\langle \langle c_{\triangle }^{},c_{\triangle }^{\dagger
}\rangle \rangle $ denotes the retarded Green function \textit{in the absence%
} of the coupling with $C_{0}$, then $\Sigma _{\triangle }(\omega )=\omega
-g_{\triangle }^{-1}(\omega )$. The Dyson equation, $\bm{\mathcal{G}}=\bm{g}+%
\bm{g}\,\bm{V}\,\bm{\mathcal{G}}$, relates the unperturbed Green function $%
\bm{g}$ with the Green funtion $\bm{\mathcal{G}}$ in the presence of the
perturbation $\bm{V}$. Taking the hopping between the site ``$0$" and its three
nearest-neighbors as the perturbation, we can immediately obtain the
following expression for the Green function of the site ``$0$" (corresponding to the $p_z$-orbital of C$_0$), 
\begin{equation}
\mathcal{G}_{0}=g_{0}+g_{0}\,3t^{2}\,\mathcal{G}_{0}\,g_{\triangle }
\end{equation}
where $g_{0}(\omega )=(\omega +\mathrm{i}0^{+})^{-1}$ and\footnote{%
Here, it is understood that $\mathcal{G}_{0}$ is evaluated with an
in\-fi\-ni\-te\-si\-mal imaginary part, $\mathcal{G}_{0}(\omega +\mathrm{i}%
0^{+})$} 
\begin{equation}
\mathcal{G}_{0}(\omega )=\frac{\omega }{2\pi t^{2}}\,\mathcal{S}\left( \frac{%
\omega ^{2}-3t}{2t}\right) \mathcal{Q}\left( \frac{\omega ^{2}-3t}{2t}\right)
\end{equation}%
with 
\begin{equation}
\mathcal{S}(x)\!=\!8(\sqrt{2x+3}-1)^{-\frac{3}{2}}(\sqrt{2x+3}+3)^{-\frac{1}{%
2}}\,.
\end{equation}%
and 
\begin{equation}
\mathcal{Q}(x)\!=\!\left\{ 
\begin{tabular}{l}
$K(k(x)^{2})$ \qquad if $\mathrm{Im}(x)\,\mathrm{Im}(k)\!<\!0$ \\ 
$K(k(x)^{2})\!+\!2\mathrm{i}\,p(x)K(1\!-\!k(x)^{2})$ otherwise%
\end{tabular}%
\ \right.
\end{equation}%
with $K(k)$ the complete elliptic integral of the first kind, $p(x)=\mathrm{%
sign}(\mathrm{Im}(x))$ and $k(x)=(2x+3)^{1/4}\mathcal{S}(x)/2$. After some
straightforward algebra we finally get 
\begin{equation}
\Sigma _{\triangle }(\omega )=\omega -\frac{3t^{2}}{\Sigma _{0}(\omega )}
\end{equation}%
where $\Sigma _{0}(\omega )=\omega -\mathcal{G}_{0}^{-1}(\omega )$. We
notice that, in the low energy limit ($|\omega| \rightarrow 0$), 
\begin{equation}
-\mathrm{Im}(\Sigma _{\triangle }(\omega ))\simeq -3t^{2}\,\mathrm{Im}(%
\mathcal{G}_{0}(\omega ))=\sqrt{3}\,|\omega |\,.
\end{equation}%
This shows that the effective density of states of the part of the graphene
sheet that couples to the symmetric state defined above also presents a
pseudo-gap and goes linearly to zero [see  Fig. \ref{sigmas}(a)].

A similar analysis can be done for the green function and the self-energy corresponding to the $|\pm\rangle$ states, $g_\pm(\omega)$ and $\Sigma_\pm(\omega)$, respectively.
We then obtain
\begin{equation}
g_\pm(\omega)=\frac{3}{2} \mathcal{G}_0(\omega)-\frac{1}{2}\mathcal{G}_\triangle(\omega)
\end{equation}
and
\begin{equation}
\Sigma_\pm(\omega)=\omega-g_\pm^{-1}(\omega)
\end{equation}
with $\mathcal{G}_\triangle(\omega)=[\omega-\Sigma_\triangle(\omega)-3t^2g_0(\omega)]^{-1}$.
This functions are plotted in Figs. \ref{sigmas}(b) and \ref{sigmas}(d).

From these results, it is clear that, at low energies, the $|\triangle\rangle$ state is only weakly coupled to the rest of the graphene sheet (excluding C$_0$) while the opposite is true for the other  two orthogonal states $|\pm\rangle$. This justifies our simplified model described in Section \ref{HF}.

\bibliographystyle{apsrev4-1}


%\bibliography{graphene}

\end{document}